\documentclass[prl,twocolumn,showpacs,preprintnumbers,amsmath,amssymb,superscriptaddress,floatfix]{revtex4}
\usepackage{graphicx}
\usepackage{dcolumn}
\usepackage{bm}
\usepackage{epsf}
\usepackage{color}

\begin{document}

\preprint{preprint}

\title{Electronic Structure and Magnetic Properties of the Half-metallic Ferrimagnet Mn$_{2}$VAl Probed by Soft X-ray Spectroscopies}


\author{K. Nagai}
\author{H. Fujiwara}
\email{fujiwara@mp.es.osaka-u.ac.jp}
\author{H. Aratani}
\author{S. Fujioka}
\author{H. Yomosa}
\author{Y. Nakatani}
\author{T. Kiss}
\author{A. Sekiyama} 
\affiliation{Graduate School of Engineering Science, Osaka University, Toyonaka, Osaka 560-8531, Japan}
\author{F. Kuroda}
\affiliation{Department of Quantum Matter, ADSM, Hiroshima University, Higashi-hiroshima, Hiroshima 739-8530, Japan}
\affiliation{CMI$^2$-MaDIS, National Institute for Materials Science, 1-2-1 Sengen, Tsukuba, Ibaraki 305-0047, Japan}
\affiliation{Institution of Scientific and Industrial Research, Osaka University, Ibaraki, Osaka 567-0047, Japan}
\author{H. Fujii}
\author{T. Oguchi}
\affiliation{CMI$^2$-MaDIS, National Institute for Materials Science, 1-2-1 Sengen, Tsukuba, Ibaraki 305-0047, Japan}
\affiliation{Institution of Scientific and Industrial Research, Osaka University, Ibaraki, Osaka 567-0047, Japan}
\author{A. Tanaka}
\affiliation{Department of Quantum Matter, ADSM, Hiroshima University, Higashi-hiroshima, Hiroshima 739-8530, Japan}
\author{J. Miyawaki}
\author{Y. Harada}
\affiliation{Institute for Solid State Physics (ISSP), University of Tokyo, Kashiwanoha, Chiba 277-8581, Japan}
\affiliation{Synchrotron Radiation Research Organization, University of Tokyo, Sayo-cho, Sayo, Hyogo 679-5198, Japan}
\author{Y. Takeda}
\author{Y. Saitoh}
\affiliation{Materials Sciences Research Center, Japan Atomic Energy Agency (JAEA), Sayo, Hyogo 679-5148, Japan}
\author{S. Suga} 
\affiliation{Institution of Scientific and Industrial Research, Osaka University, Ibaraki, Osaka 567-0047, Japan}
\author{R. Y. Umetsu} 
\affiliation{Institute for Materials Research, Tohoku University, 2-1-1 Katahira, Sendai 980-8577, Japan}

\date{\today}

\begin{abstract}
We have studied the electronic structure of ferrimagnetic Mn$_2$VAl single crystals by means of soft X-ray absorption spectroscopy (XAS), X-ray absorption magnetic circular dichroism (XMCD) and resonant soft X-ray inelastic scattering (RIXS). We have successfully observed the XMCD signals for all the constitute elements. The Mn $L_{2,3}$ XAS and XMCD spectra are reproduced by spectral simulations based on density-functional theory, indicating the itinerant character of the Mn $3d$ states. On the other hand, the V $3d$ electrons are rather localized since the ionic model can qualitatively explain the V $L_{2,3}$ XAS and XMCD spectra. This picture is consistent with local $dd$ excitations revealed by the V $L_3$ RIXS. 
\end{abstract}





\maketitle

\section{INTRODUCTION}

Half-metals are characterized by a peculiar electronic structure as one of the spin sub-bands is metallic and the other is semiconducting with a gap at the Fermi level ($E_\textrm{F}$). The expected 100\% spin-polarization at $E_\textrm{F}$ is suitable for functional spintronic applications. In particular, full-Heusler alloys with the chemical formula $X_2YZ$ ($X$ and $Y$: transition metals, $Z$: main group element) have been extensively studied since the Currie temperature ($T_\textrm{C}$) is , for example, as high as 985 K for Co$_2$MnSi~\cite{I. Galanakis_2016,P. J. Brown2000}.

L2$_1$ ordered Mn$_2$VAl is a ferrimagnet with $T_\textrm{C}\sim$760 K\cite{Y. Yoshida1981,R. Y. Umetsu2015}, and is predicted to be half-metallic~\cite{S. Ishida1984,R. Weht1999}. The saturation magnetic moment ($M_s$) is reported as 1.94 $\mu_\textrm{B}$ per formula unit (f.u.) for poly crystalline samples~\cite{C. Jiang2001, R. Y. Umetsu2015}, which is much smaller than that of 4.78 $\mu_\textrm{B}$/f.u. for Co$_2$MnSi~\cite{L. Ritchie2003}. The small $M_s$ due to the antiparallel spin coupling of V and Mn offers an advantage for saving the electric power of the current-induced magnetization switching, since the switching current is proportional to $M_s^2$~\cite{H.Itoh1983, J. Z. Sun2000}.

To reveal the detailed magnetic properties and the electronic structure of the Heusler alloys, intensive studies have been performed by means of X-ray absorption spectroscopy (XAS) and X-ray absorption magnetic circular dichroism (XMCD)~\cite{K. Miyamoto2003, N. D. Telling2006,N. D. Telling2008,P. Klaer2009, P. Klaer2010, M. Tsunekawa2014, G. H. Fecher2014, A. Yamasaki2002, M. Jourdan2014}. The electronic structure of Co$_2$MnSi has been known as the Co $3d$ electrons are itinerant, while the Mn $3d$ electrons contribute to the local magnetic moment~\cite{N. D. Telling2006,N. D. Telling2008,G. H. Fecher2014}. Nevertheless, only few works have been performed on Mn$_2$VAl~\cite{T. Kubota2009, M. Meinert2011, J. Karel2015} because of the difficulty in synthesizing high quality crystals. It has been reported as the epitaxially grown Mn$_2$VAl films are prone to various disorder~\cite{T. Kubota2009}. Since the disorder effects change the magnetic properties and the local electronic structure~\cite{M. Meinert2011}, the line shapes of XAS and XMCD spectra varied depending on the individual samples~\cite{T. Kubota2009, M. Meinert2011, J. Karel2015}. Therefore, the electronic structure of Mn$_2$VAl is still controversial.

In this work, we present the detailed electronic structure and magnetic properties of the high quality single crystals of L2$_1$-ordered Mn$_2$VAl, revealed by means of XAS and XMCD for all constitute elements. The magnetic moments of the Mn sites evaluated by using the magneto-optical sum-rule~\cite{B. T. Thole1992,P. Carra1993} were qualitatively consistent with the results of density functional theory (DFT) within the experimental accuracy. We found that the line shapes of Mn $L_{2,3}$ XAS and XMCD spectra of Mn$_2$VAl were deviated from those of Mn spectra in Co$_2$MnSi, and are well explicable  by the spectral simulation based on DFT, indicating the itinerant character of the Mn $3d$ electrons. On the other hand, the V $L_{2,3}$ XAS and XMCD spectra were qualitatively described in the ionic model.  Moreover, the $dd$ excitation in the V $3d$ states was probed using resonant soft-X-ray inelastic scattering (RIXS)~\cite{Book_Kotani_deGroot,Ament_Review_RMP}, suggesting that the V $3d$ electrons are localized.

\section{Experimental}
\subsection{Characterization of single crystals}

Polycrystalline mother ingots of Mn$_2$VAl and Co$_2$MnSi were prepared by induction melting in an argon gas atmosphere. Since the melting temperature of V is very high as $\sim$2180 K and the vapor pressure of V is high enough during the induction melting process, the excess Mn ions are contained in the mother ingot of Mn$_2$VAl. Single crystals of both Mn$_2$VAl and Co$_2$MnSi were grown by the Bridgeman method with a size of 12 mm in diameter and about 30 mm in length. Obtained ingots were annealed at 1473 (1373) K in the case of Mn$_2$VAl (Co$_2$MnSi) to gain the size of single crystal grains. For Mn$_2$VAl, a two-step annealing process at 1123 K and 873 K was employed and the crystal was finally slow-cooled to room temperature to control the degree of order. The specimen of Co$_2$MnSi was obtained by slow cooling from 1373 K to room temperature. Crystal orientation was checked with a Laue camera and the specimen was cut in a strip form in the direction parallel to $<$100$>$ for Mn$_2$VAl and $<$110$>$ for Co$_2$MnSi. Composition of the specimens was confirmed with an electron probe microanalyzer (EPMA) to be Mn: 50.5, V: 26.9, Al: 22.6 for Mn$_2$VAl and Co: 50.0, Mn: 25.9, Si: 24.1 for Co$_2$MnSi. Moreover, the long-range order parameter for the L2$_1$ structure (S$_{\rm{L}2_1}$)~\cite{Webster1971,Takamura2009} was evaluated as 0.84 for Mn$_2$VAl and 0.90 for Co$_2$MnSi by using the X-ray diffraction measurements. Note that the S$_{\rm{L}2_1}$ value for Mn$_2$VAl was much higher than that for the epitaxial films of S${_{\rm{L}2_1}}\sim0.5$\cite{T. Kubota2009, M. Meinert2011}, indicating the well ordered single crystal. 

The bulk magnetization was measured with a superconducting quantum interface device (SQUID) magnetometer. We obtained a total magnetic moment of $m_\textrm{SQUID}$ = 1.82 $\mu_\textrm{B}$/f.u. for Mn$_2$VAl single crystals, which is slightly smaller than the value of 1.94 $\mu_\textrm{B}$/f.u for polycrystalline samples~\cite{C. Jiang2001, R. Y. Umetsu2015} and the expected value of 2 $\mu_\textrm{B}$/f.u. from the Slater-Pauling rule~\cite{I. Galanakis_2016}. This deviation would be due to off-stoichiometric effects; the small amount of excess Mn and V ions substituted into Al sites can reduce the total magnetic moment, (see APPENDIX).

\subsection{Soft X-ray spectroscopies}
Single crystalline samples were fractured \textit{in situ} in ultrahigh vacuum to obtain the clean surface. XAS and XMCD measurements were performed at BL23SU in SPring-8~\cite{Y. Saitoh2012}. The spectra were recorded in total-electron-yield mode with an energy resolution better than 0.1 eV using a superconducting magnet in fields up to 2 T along the incident beam direction. To eliminate any experimental artifacts arising from system errors, each XMCD spectrum was measured for opposite orientations of the applied magnetic field and the resulting spectra were averaged. The measurement temperature was set to 20 K. The RIXS measurements were performed at the SPring-8 BL07LSU \textit{`HORNET'} end-station~\cite{Harada2012,Yamamoto2014,Miyawaki2017}. The total energy resolution was set to $\sim$200 meV at the measurement temperature of 300 K.

\section{Density Functional Theory}

The electronic structure calculation based on DFT has been performed using the HiLAPW code, which is based on the all-electron full-potential augmented plane-wave (FLAPW) method\cite{E. Wimmer1981}. The generalized gradient approximation (GGA) using the Perdew-Burke-Ernzerhof scheme has been used for the exchange correlation potential\cite{J. P. Perdew1996, J. P. Perdew1997}. We have used the second-variation procedure to include the spin-orbit coupling in addition to a scalar-relativistic scheme, giving an accurate description equivalent to solving the Dirac equation for relatively light elements as $3d$ transition metals. Plane-wave expansion cutoffs were set to 20 Ry for wave functions and 80 Ry for charge density and potential functions. The muffin-tin sphere radius was chosen as 1.1 {\AA} for all elements. For the Brillouin-zone integration, a 16$\times$16$\times$16 mesh was used with the tetrahedron integration technique. The atoms were placed on the general form $X_2YZ$ of the L$2_1$ structure with $X$ at ($\frac{1}{4}$,$\frac{1}{4}$,$\frac{1}{4}$) and ($\frac{3}{4}$,$\frac{3}{4}$,$\frac{3}{4}$), $Y$ at (0,0,0), and $Z$ at ($\frac{1}{2}$,$\frac{1}{2}$,$\frac{1}{2}$). The lattice constant was set to 5.875 {\AA}~\cite{R. Weht1999}.

\begin{figure}[htbp]
 \begin{minipage}{1\hsize}
 \begin{center}
 \includegraphics[width=80mm,clip]{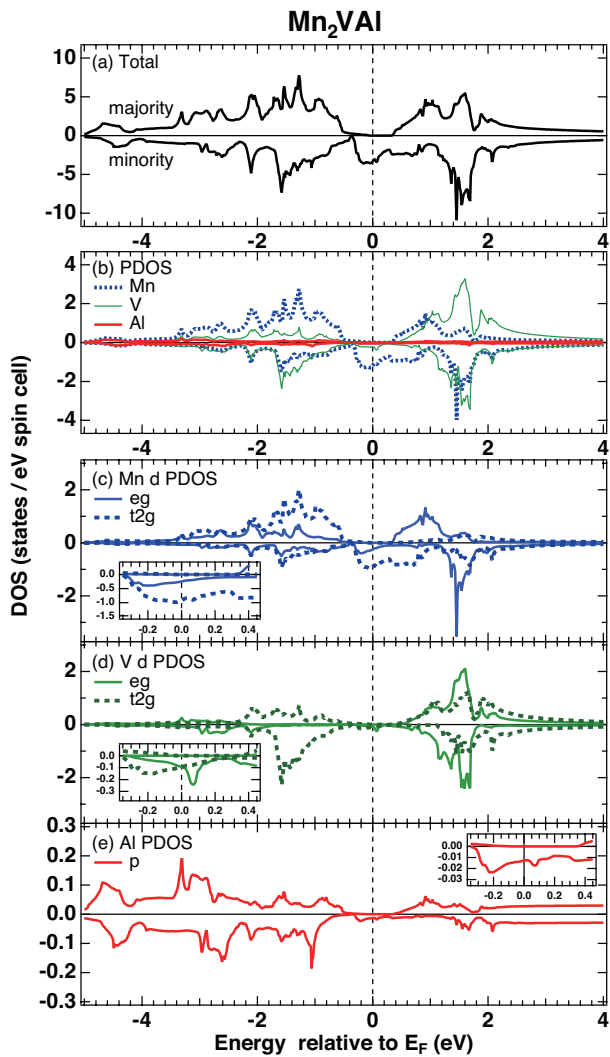}
 \end{center}
 \end{minipage}
 \caption
 {(Color online) (a) Total density of states of L2$_1$ ordered Mn$_2$VAl. (b) PDOS of Mn, V, Al sites. (c), (d) PDOS of Mn and V $3d$ states projected onto the $e_g$ and $t_{2g}$ orbital. (e) PDOS of Al $3p$ states. The insets show the enlarged view of PDOS around $E_\textrm{F}$.}
  \label{DOS}
 \end{figure}

Figure \ref{DOS}(a) shows the calculated spin-polarized density of states for Mn$_2$VAl with an half-metallic gap for the majority-spin states, which is qualitatively consistent with previous reports~\cite{S. Ishida1984,R. Weht1999}. The partial density of states (PDOS) in Figs. \ref{DOS}(b) and \ref{DOS}(c) show that the Mn $3d$ minority-spin sub-bands contribute mainly to the states in the vicinity of $E_\textrm{F}$. Meanwhile, PDOS of the V $3d$ states are only $\sim$10\% of PDOS of the Mn $3d$ states around $E_\textrm{F}$ as shown in the insets. Clear splitting of the V $3d$ states into the occupied $t_{2g}$ and unoccupied $e_{g}$ states is seen with the band splitting energy of $\sim$3.2 eV as shown in Fig. \ref{DOS}(d). The PDOS of the Al $3p$ states are even smaller and $\sim$10\% of the PDOS of the V $3d$ states at $E_\textrm{F}$.

The whole XMCD spectra were simulated within the dipole approximation on the basis of the FLAPW method, where the transition matrix elements were calculated only within the muffin-tin sphere. The core-hole potential in the final states was neglected in the present calculations. To take into account the experimental core-hole lifetime broadening, the calculated spectra were broadened using Lorentzian functions with $\gamma$ =  0.97(0.36) eV and 0.78(0.28) eV for the $L_2(L_3)$ edges of Mn and V, respectively, while 0.15 eV was employed for the Al $K$ edge. The detailed information about the calculation has been given in Ref.~\onlinecite{H. Fujii2014}.

\section{RESULTS AND DISCUSSIONS}
\subsection{Magnetic properties probed by XMCD and sum-rule analysis}

\begin{figure*}[htbp] 
  \begin{center}
       \includegraphics[width=160mm]{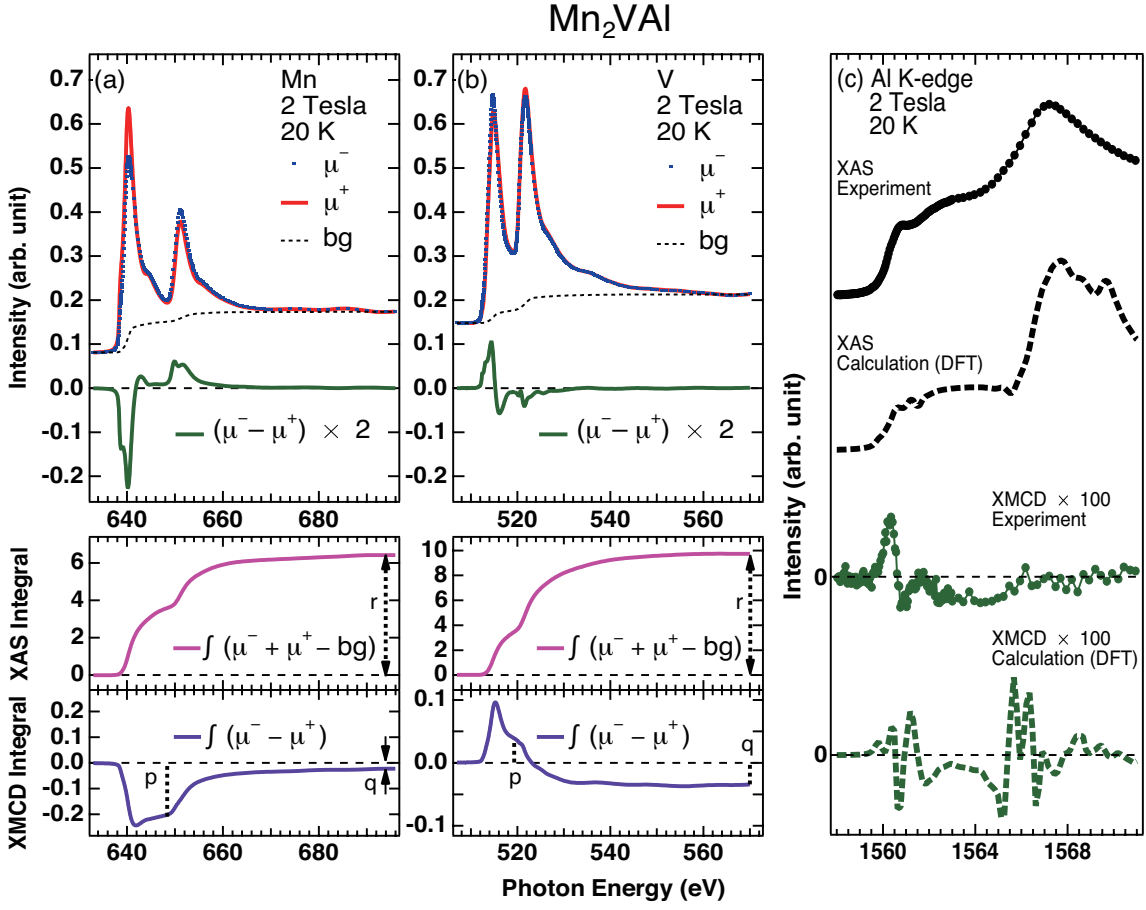}
    \end{center}
 \caption
 {(Color online) XAS and XMCD spectra of Mn${_2}$VAl recorded at T = 20 K with H = 2 T. (a) Mn and (b) V $L_{2,3}$ XAS and XMCD spectra (top), together with the integrated curves of XAS (middle) and XMCD (bottom). (c) Al $K$ XAS and XMCD spectra with the DFT-based simulation.} 
   \label{Mn_V_MVA}
\end{figure*}

Figures \ref{Mn_V_MVA}(a) and \ref{Mn_V_MVA}(b) show the experimental results of Mn and V $L_{2,3}$ XAS and XMCD spectra of Mn$_2$VAl. $\mu$$^+$ ($\mu$$^-$) denotes the XAS absorption intensity for parallel (antiparallel) alignment of the photon helicity and sample magnetization direction. The XAS spectra are normalized by the $L_{3}$ peak intensity of the polarization-summed XAS (${\mu^-}+{\mu^+}$). There are fine structures in the XMCD (${\mu^-} - {\mu^+}$) spectra for both Mn and V $L_{2,3}$ edges, and the sign of the XMCD signals is opposite between them, reflecting ferrimagnetic coupling between the Mn and V spins (Mn along applied field). The detailed line shape is discussed in the next subsection (I$\hspace{-.1em}$V. B).

Moreover, we can clearly see sizable XMCD signals even in the Al $K$ edge as shown in Fig. \ref{Mn_V_MVA}(c) despite its non-magnetic atomic character~\cite{J. Schmalhorst2007, R. Y. Umetsu2010}, indicating a non negligible orbital polarization 
of the valence states with Al $3p$ character. The most prominent XMCD peak is located just at the threshold. The XMCD signals originate from a combination of the spin-orbit and exchange splittings of the Al $3p$ states.  The integration of the XMCD signals over the K edge, which is proportional to the orbital magnetic moment~\cite{G. Y. Guo1996,Book_Kotani_deGroot}, falls to zero within the experimental accuracy.

The theoretical simulations are consistent with the experimental spectra, although there are some discrepancies in XMCD. The DFT calculation yields an orbital moment of $m_{L} = 0.001$ $\mu_\textrm{B}$/Al with an orbital-to-spin magnetic moment ratio of $m_{L}/m_{S}=-0.05$. The deviation between the experiment and theory especially on the higher energy region is possibly due to the energy dependent lifetime broadening effects, which are not taken into account in the simulation, since the broadening width is given by fitting the edge step of the XAS spectrum.


To reveal more detailed magnetic properties of the transition metal sites, the orbital and spin magnetic moments per $d$ hole were estimated by applying the magneto-optical sum-rule~\cite{B. T. Thole1992, P. Carra1993,VanDerLaanReview,Edmonds2015} using the equations,
\begin{eqnarray}
\frac{m_{spin}}{n_h} = -\frac{6p-4q}{r{P_{c}}}C,\\
\frac{m_{orb}}{n_h} = -\frac{4q}{3r{P_{c}}},\ \ \ \ \ \ \   
\end{eqnarray}
where $n_h$ is the number of unoccupied $d$ holes, $p$ ($q$) is the integral of the XMCD signal over the $L_3$($L_{2, 3}$)-edge, and $r$ is the integral of the polarization summed XAS intensity after subtracting the background. As shown in the bottom of Figs. \ref{Mn_V_MVA}(a) and \ref{Mn_V_MVA}(b), the cut off energy to obtain $p$ was set to the minimum of the polarization summed XAS intensity between the $L_2$ and $L_3$-edges. The $q$ and $r$ values were evaluated at the energy of 694 eV for Mn and 570 eV for V. $P_{c}$ denotes the degree of circular polarization of 0.97~\cite{Y. Saitoh2012}. 
In addition, we must take into account the so-called $jj$ mixing effect by using the correction factor C for the spin sum-rule~\cite{E. Goering2005}. Assuming C = 1.5 for Mn\cite{H. A. Durr1997}, we obtained a spin (orbital) magnetic moment of $m_{spin}^{Mn}$($m_{orb}^{Mn})$  = 0.27 (0.005) $\mu_{B}$/Mn/hole, which is consistent with that obtained with our DFT calculation as summarized in Table \ref{sumrule}. Assuming $n_h$ = 5.20 estimated from the DFT, we obtained a total Mn moment of 2.86 $\mu_\textrm{B}$/f.u.


\begin{table} [hbt]
\begin{center}
\caption{Spin and orbital magnetic moments of Mn$_2$VAl in unit of $\mu_\textrm{B}$ per hole for an Mn (V) site obtained by the sum-rule analysis and DFT.} 
\begin{tabular}{ccccccc}
Mn$_2$VAl &&&&&&\\
\hline
\hline
& $m_{spin}$/$n_h$ &\  $m_{spin}$ /$n_h$   &  \   $m_{orb}$/$n_h$ & \   $m_{orb}$/$n_h$   \\
 &\ (XMCD) &\ (DFT)&\    (XMCD)&   \ (DFT)    &\\       
\hline
Mn& 0.27 & 0.28 & \     0.005 &\     0.006  & \\
V   &   -0.15  &  -0.11  & \    0.005  &  \  0.001 &\\
\hline
\hline


\end{tabular}
  \label{sumrule}
 \end{center}
 \end{table}
 
Using the orbital sum-rule for V, we obtained an orbital magnetic moment of $m_{orb}^{V}$ = 0.005 $\mu_{B}$/V/hole as given in Table I. Meanwhile, it is not straight forward to adopt the spin sum-rule for V since the C value is not established and ranges from 3.6 to 5.0 due to the small spin-orbit splitting at the V $L_{2,3}$ edges~\cite{E. Goering2005,A. Scherz2002}. If we evaluate $m_{spin}^{V}$ using the spin sum-rule assuming C = 1, we obtained $m_{spin}^{V}$ = -0.04 $\mu_\textrm{B}$/V/hole. In this case, we obtained a total V moment of -0.26 $\mu_\textrm{B}$/f.u., assuming $n_h$ =7.39 estimated from the DFT. On the other hand, we can evaluate the total V moment as $m^{V} = -1.04$ $\mu_\textrm{B}$/f.u. using the total Mn moment (2.86 $\mu_\textrm{B}$/f.u.) and the total bulk moment (1.82 $\mu_\textrm{B}$/f.u.) obtained by SQUID. Therefore, we found that a likely correction factor for V is C = 3.8, being comparable to that of 3.6 given in Ref. \onlinecite{E. Goering2005}. The spin (orbital) moment for V is deviated from the DFT value of $m_{spin}^{V}$($m_{orb}^{V}$) = -0.11 (0.001) $\mu_\textrm{B}$/V/hole as listed in Table \ref{sumrule}, implying the limitation of DFT for describing the electronic structure of the V $3d$ states as discussed later.

\begin{figure}[htbp]
 \begin{minipage}{1\hsize}
  \begin{center}
   \includegraphics[width=80mm]{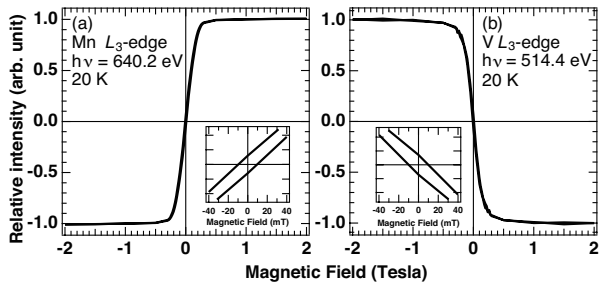} 
  \end{center}
 \end{minipage}
 \caption
 {(Color online) (a), (b) Magnetic field dependence of the XMCD intensity at the Mn and V $L_3$ peak. The inset shows the expanded view around the 0 T.} 
   \label{MH_MnV}
\end{figure}

Figure \ref{MH_MnV} (a) shows the relative intensity of the Mn $L_3$ XMCD signal at 640.2 eV on sweeping the field from -2 to 2 T, giving the element specific magnetization profile of the Mn sites. Meanwhile, the field dependence of the V $L_3$ XMCD intensity at 514.4 eV shows the opposite trend reflecting the antiferromagnetic spin coupling to the Mn moment as shown in Fig. \ref{MH_MnV}(b). The enlarged views around 0 T in the insets of Figs. \ref{MH_MnV}(a) and \ref{MH_MnV}(b) give a coercivity of 10 mT for both Mn and V sites, which is about 10\% for the epitaxial films~\cite{T. Kubota2009}. Moreover, the XMCD intensity does not change after saturating the magnetization, and thus the paramagnetic component reported in Ref.~\onlinecite{T. Kubota2009} is negligible. Therefore we stress that the impurity and disorder effects are successfully suppressed in our single crystalline samples.

\subsection{XAS and XMCD line shapes: itinerant v.s. localized character in the electronic states}

\begin{figure}[htbp]
 \begin{minipage}{1\hsize}
  \begin{center}
   \includegraphics[width=65mm]{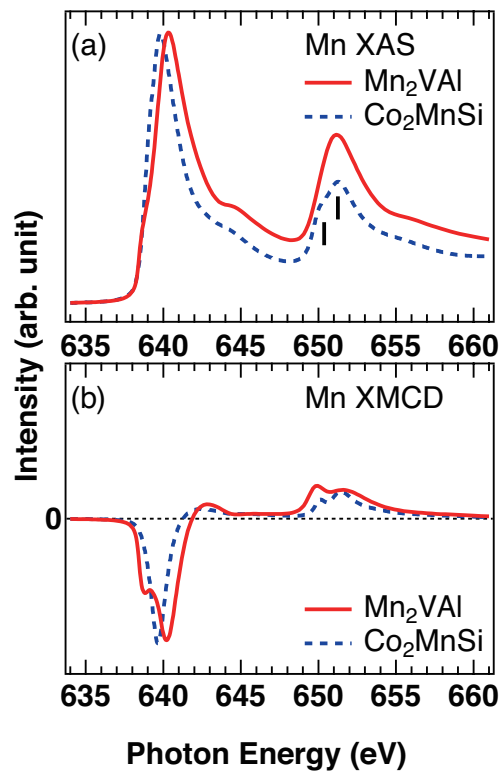}
  \end{center}
 \end{minipage}
 \caption
 {(Color online) Mn $L_{2, 3}$-edges XAS (a) and XMCD (b) spectra for Mn$_2$VAl(solid line) and Co$_2$MnSi(dashed line).} 
\label{compare_MVA_CMS}
\end{figure}

To discuss the characteristic electronic properties of Mn$_2$VAl, we focus on the detailed line shapes of the polarization-summed XAS and XMCD spectra for both Mn and V $L_{2,3}$ edges. Figure~\ref{compare_MVA_CMS}(a) displays the enlarged view of the Mn $L_{2,3}$ XAS spectrum of Mn$_2$VAl, showing a metallic line shape as reported in Ref.~\onlinecite{J. Karel2015}. Compared with the XAS spectrum of Co$_2$MnSi, in which the Mn $3d$ states have a localized $3d^5$ character~\cite{N. D. Telling2006,N. D. Telling2008,G. H. Fecher2014}, the $L_3$ main line of Mn$_2$VAl shifts to higher energies even though the leading edges start at the same energy of 638 eV. Moreover, the $L_2$ peak of Mn$_2$VAl does not show the doublet structure due to the atomic multiplets observed in Co$_2$MnSi~\cite{K. Miyamoto2003,N. D. Telling2006,N. D. Telling2008,G. H. Fecher2014,A. Kimura1997, B. T. Thole1985}.

In contrast, the XMCD spectrum of Mn$_2$VAl shows clear doublet peaks in the $L_2$ edge as well as a clear shoulder structure on the low energy side of the $L_3$ main peak as shown in Fig.~\ref{compare_MVA_CMS}(b). This line shape is qualitatively consistent with that reported in Ref. \onlinecite{M. Meinert2011}. Note that the Mn $L_3$ XMCD spectrum of Co$_2$MnSi shows the single peak located at the middle of the $L_3$ main line in contrast to the low-energy shoulder present in Mn$_2$VAl, illustrating a clear difference in the local electronic structure on the Mn sites between these two materials.
 
\begin{figure}[htbp]
 \begin{minipage}{1\hsize}
  \begin{center}
   \includegraphics[width=65mm]{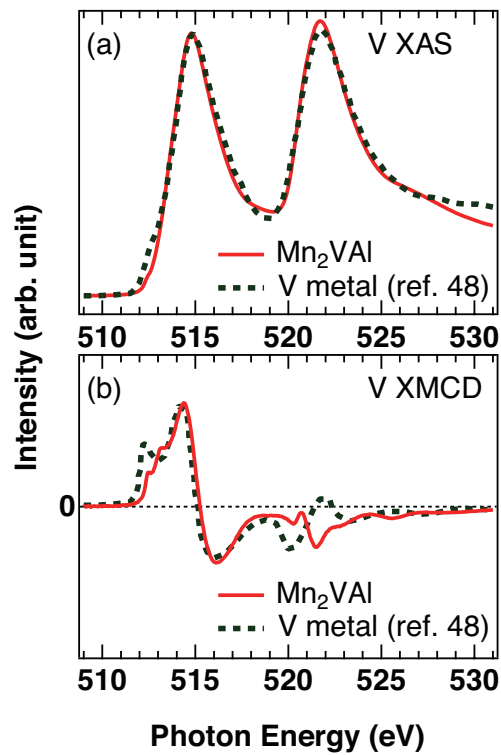}
  \end{center}
 \end{minipage}
 \caption
 {(Color online) V $L_{2, 3}$-edges XAS (a) and XMCD (b) spectra for Mn$_2$VAl(solid line) and V metal (dashed line) reported in the reference \cite{H. Wende2004}. The XMCD spectra are normalized at the $L_{3}$ peak.}
   \label{compare_MVA_V}
\end{figure}

In Fig.~\ref{compare_MVA_V} we show the V  $L_{2,3}$ XAS and XMCD spectra of Mn$_2$VAl together with those of V metal reported in Refs.~\onlinecite{A. Scherz2002} and \onlinecite{H. Wende2004} to discuss the electronic structure of the $Y$-site element. The V  $L_{2,3}$ XAS spectrum of Mn$_2$VAl is qualitatively similar to that for V metal. Nevertheless, the XMCD spectrum of Mn$_2$VAl significantly deviates from that of V metal. The single XMCD peak before the $L_3$ main line observed in V metal is suppressed, while two tiny shoulder structures show up before the $L_3$ main peak in Mn$_2$VAl~\cite{T. Kubota2009, M. Meinert2011}. In addition, the $L_2$ XMCD shape for Mn$_2$VAl is clearly different from that of V metal, which is well reproduced by DFT-based simulations~\cite{A. Scherz2002,H. Wende2004}. These results suggest the localized character of the V $3d$ states in Mn$_2$VAl.

\begin{figure}[htbp]
 \begin{minipage}{1\hsize}
  \begin{center}
\includegraphics[width=60mm,clip]{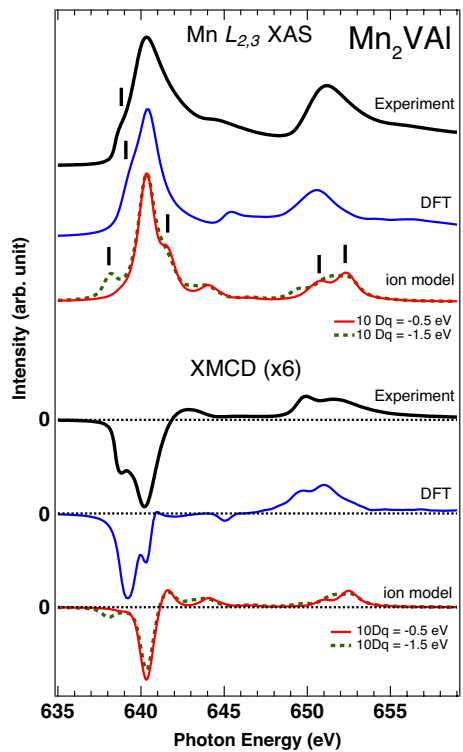}
  \end{center}
 \end{minipage}
  \caption
 {(Color online) Mn $L_{2, 3}$-edge XAS and XMCD spectra compared with the simulation based on DFT and the ionic model. Spectra are normalized by the $L_3$-peak intensity of the polarization summed XAS spectra.}
\label{calc_Mn}
\end{figure}

To obtain further insight of the electronic structure for the Mn and V $3d$ states, we have analyzed the line shapes of XAS and XMCD spectra using simulations based on the DFT and the ionic model, representing the itinerant and localized limit of the $3d$ states. The ionic calculations based on the full multiplet theory were implemented using the XTLS 9.0 program~\cite{A. Tanaka1994}. The local crystalline electric field (CEF) was taken into account for the Mn$^{2+}$ ion with tetrahedral ($T_d$) symmetry and the V$^{2+}$ ion with octahedral ($O_h$) symmetry. All the intra-atomic parameters such as the $3d$-$3d$ and $2p$-$3d$ Coulomb and exchange interactions (Slater integrals) and the $2p$ and $3d$ spin-orbit couplings have been obtained using Cowan's Hartree-Fock code~\cite{R. D. Cowan1981} and reducing the Slater parameters to 80\%.

\begin{figure}[htbp]
 \begin{minipage}{1\hsize}
  \begin{center}
\includegraphics[width=60mm,clip]{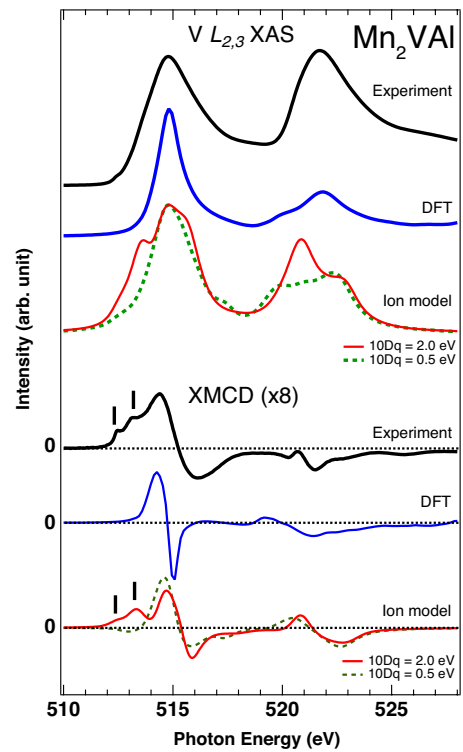}
  \end{center}
 \end{minipage} 
 \caption
 {(Color online) V $L_{2, 3}$-edge XAS and XMCD spectra compared with the simulation based on DFT and the ionic model. Spectra are normalized by the $L_3$-peak intensity of the polarization summed XAS spectra.}
   \label{calc_V}
\end{figure}

In Fig.~\ref{calc_Mn}, we show the results of spectral simulations for the Mn $L_{2,3}$ XAS and XMCD. The ionic model XAS calculation with 10Dq = -0.5 eV~\cite{N. D. Telling2008} shows multiplet structures with a shoulder on the higher energy side of the $L_{3}$ main line and the $L_{2}$ doublet. Moreover, an additional low-energy satellite at the $L_3$ peak appeared, reflecting the CEF splitting of 10Dq = -1.5 eV. However, such multiplet structures are not observed in the experimental XAS spectrum. Furthermore, the XMCD simulation based on the ionic model is obviously deviated from the experimental result, whereas the DFT-based simulation better reproduces the XAS and XMCD spectra, particularly for the low energy structure at the $L_{3}$ edge, indicating the itinerant character of the Mn $3d$ states in Mn$_2$VAl.

On the other hand, the situation is different in the case of V $L_{2,3}$ XAS spectra as shown in Fig.~\ref{calc_V}. The DFT-based simulation fails to explain the relative peak-intensity ratio of $L_{3}$ and $L_{2}$ XAS. Moreover, the line shape of XMCD spectrum is not explained by DFT especially for the low-energy shoulders of the $L_{3}$ XMCD. In addition, one may also notice that the $L_{2}$ XMCD is also deviated from DFT in contrast to the case of V-metal~\cite{A. Scherz2002,H. Wende2004} (Fig.~\ref{compare_MVA_V}(b)), implying the limitation of the band picture for the V $3d$ states in the case of Mn$_2$VAl. Meanwhile, the ionic calculation reproduces the experimental results rather well with a CEF parameter of 10Dq = 2 eV  for $O_h$ symmetry, which is not dramatically smaller than the band-splitting energy between the V  $3d$ $t_{2g}$ and $e_{g}$ states obtained by the DFT as shown in Fig~\ref{DOS}. These results suggest that the V $3d$ electrons are rather localized and thus the atomic multiplets should be taken into account for the spectral simulations.

\begin{figure}[h]
 \begin{minipage}{1\hsize}
  \begin{center}
   \includegraphics[width=80mm,clip]{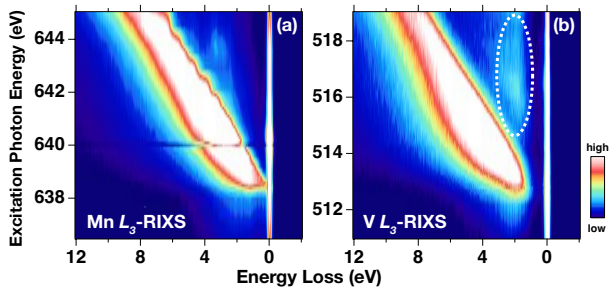}
  \end{center}
 \end{minipage}
 \caption
 {(Color online) Intensity plots of $h\nu_{in}$-dependent RIXS at the Mn (a) and V (b) $L$-edges. Dashed circle is the guide for the eye indicating the resonant excitation.}   
\label{rixs}
\end{figure}

To check the localized nature of the $3d$ electrons, we have performed Mn and V $L_3$ RIXS, which is a powerful tool to probe the local electron excitation in the $3d$ states~\cite{Ghiringhelli2006,Benckiser2013}. As shown in Fig.~\ref{rixs}(a), the energy-dispersive fluorescence components are dominant in Mn $L_3$ RIXS, supporting the itinerant picture~\cite{Yang2009} for the Mn $3d$ states. Meanwhile, the V $L_3$ RIXS in Fig.~\ref{rixs}(b) clearly shows the clear resonant inelastic excitation ($dd$ excitation) with the constant energy-loss component around 2 eV, which is consistent with the 10Dq value of ionic calculations for the V $L_{2,3}$ XAS and XMCD spectra. Therefore, we stress that this $dd$ excitation indicates the localized character of the V $3d$ electrons in Mn$_2$VAl. 

\section{CONCLUSIONS}

We investigated the element specific magnetic properties and electronic structure of three different constitute elements of high quality single crystal Mn$_2$VAl by using XAS and XMCD. The successful observation of the Al $K$ XMCD signals suggests the very small spin unbalance of the Al $p$ states near $E_\textrm{F}$ induced by the hybridization with the transition metal $3d$ states. The Mn and V $L_{2,3}$ edge XMCD spectra show the ferrimagnetic spin coupling between Mn and V states, and the magnetic moments evaluated by the sum-rule analysis has shown an excellent agreement with DFT calculation for its Mn sites. The electronic structure analysis for the Mn $L_{2,3}$ XAS and XMCD spectra with the DFT-based simulation reveals the itinerant character of the Mn $3d$ states, whereas the V $L_{2,3}$ XAS and XMCD spectra are qualitatively explained by the ionic model. In addition, V $L_{3}$ RIXS showed the $dd$ excitation, indicating the localized character of the V $3d$ states. Our results give the significant remark to design the magnetic properties by controlling the itinerant and localized character of the $d$ states depending on the site symmetry in Heusler alloys.

\section*{ACKNOWLEDGMENTS}

We thank T. Kanomata, S. Imada, A. Yamasaki, K. Yamagami, M. Kawada, A. Koide and A. Kimura for fruitful discussions. The measurements were performed under the approval of BL23SU and BL07LSU at SPring-8 (Proposal Nos. 2016A3832 (E28) and 2016B7512). FK, HF, and TO would like to thank support by ÓMaterials research by Information IntegrationÓ Initiative from CMI$^2$-MaDIS, NIMS. This work was financially supported by a Grant-in Aid for Scientific Research (JP16H04014) from the Ministry of Education, Culture, Sports, Science and Technology, Japan. A part of this work was supported by Japan Science and Technology Agency, Precursory Research for Embryonic Science and Technology (JST-PRESTO). 

\section*{APPENDIX: OFF-STOICHIOMETRY EFFECTS}
\begin{figure}[htbp]
 \begin{minipage}{1\hsize}
 \begin{center}
 \includegraphics[width=80mm,clip]{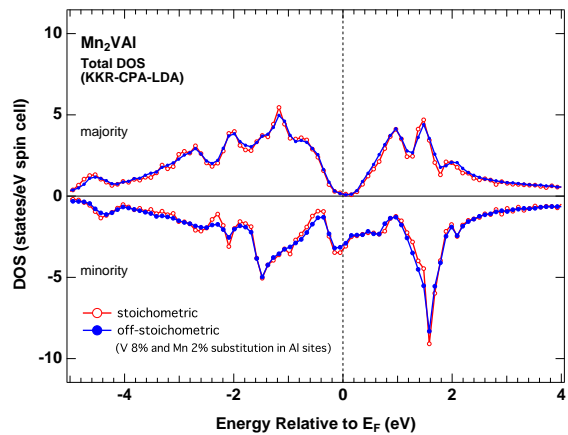}
 \end{center}
 \end{minipage}
 \caption
 {(Color online) Off-stoichiometric effects on total density of states of Mn$_2$VAl obtained by the KKR-CPA-LDA method.}
  \label{KKR-CPA}
 \end{figure}

The off-stoichiometry effects in Mn$_2$VAl were numerically simulated by the Korringa-Kohn-Rostoker (KKR) method incorporated with the coherent-potential approximation (CPA) in the local density approximation (LDA) implemented by Machikaneyama-2002 package~\cite{H.Akai1989} as demonstrated in Fig.~\ref{KKR-CPA}. The atomic positions and the lattice constant were set to the same values as Section III. Assuming that the Al sites are substituted by excess ions with 2\% for Mn and 8\% for V, which is evaluated by EPMA for the measured specimen, we obtained the total magnetic moment of 1.71 $\mu_\textrm{B}$/f.u. This value was smaller than that of the stoichiometric composition of 1.96 $\mu_\textrm{B}$/f.u. The obtained spin polarization at $E_\textrm{F}$ defined as $P = (N_{\uparrow}-N_{\downarrow})/(N_{\uparrow}+N_{\downarrow})$, where $N_{\uparrow}$ ($N_{\downarrow}$) denotes DOS at $E_\textrm{F}$ for the majority (minority) spin subband in Fig.~\ref{KKR-CPA}, is obtained as -0.93 (-0.91) for off-stoichometric (stoichometric) composition. Note that the total density of states are not significantly modified due to the off-stoichiometry effects within the simulation as shown in Fig.~\ref{KKR-CPA}. 

\end{document}